\documentclass[pre,onecolumn]{revtex4}
\usepackage{amssymb}
\usepackage{amsmath}
\usepackage{graphicx}
\begin{document}
\title{Slow decay of infection in the inhomogeneous SIR model}
\author{Hidetsugu Sakaguchi and Yuta Nakao}
\address{Department of Applied Science for Electronics and Materials,
Interdisciplinary Graduate School of Engineering Sciences, Kyushu
University, Kasuga, Fukuoka 816-8580, Japan}
\begin{abstract}
The SIR model with spatially inhomogeneous infection rate is studied with numerical simulations in one, two, and three dimensions, considering the case that the infection spreads inhomogeneously in densely populated regions or hot spots. We find that the total population of infection decays very slowly in the inhomogeneous systems in some cases, in contrast to the exponential decay of the infected population $I(t)$ in the SIR model of the ordinary differential equation. The slow decay of the infected population suggests that the infection is locally maintained for long and it is difficult for the disease to disappear completely. 
 
\end{abstract}
\maketitle
\section{Introduction}
Various nonlinear phenomena such as limit-cycle, chaos, and pattern formation have been studied with numerical simulations. 
Population dynamics is an important research field for nonlinear phenomena. 
The temporal oscillation in ecosystems owing to the prey-predator interaction is reproduced by the Lotka-Volterra equation~\cite{LV}. Chaotic dynamics was found in the logistic map  and it is applied to explain annual number fluctuations of bean weevils by May~\cite{lg}. It is important to understand the spread of diseases from population dynamics. Kermack and McKendrick proposed a mathematical model for epidemics in 1927~\cite{KM}. The model describes the population dynamics of $S$ (susceptible), $I$ (infected), and $R$ (recovered). Many authors studied the Kermack-McKendrick model and generalized models, and applied it to the analysis of the spread of diseases~\cite{Murray,Diekmann, Capasso}. Nobel performed numerical simulation of the geographic and temporal development of plagues using the mathematical model with diffusion terms~\cite{Nobel}.  The model was also used to study vaccination strategy~\cite{Shulgin,Laguzet}. The recent COVID-19 pandemic have induced intensive study of mathematical models of epidemic~\cite{Science, Costa}. 

In statistical physics, simplified stochastic models of infection on lattices called contact process have been intensively studied. It was shown that the critical behavior of the contact process is in the same universality class of directed percolation~\cite{Grassberger, Kinzel, Dickman, Henkel, Tome}. In the contact process, there is a phase transition from an infection phase to no infection phase on finite-dimensional lattices.   
Pastor-Satorras and Vesignani studied the spread of infection in scale-free networks~\cite{Ves}. They showed that the infection does not disappear even for sufficiently small infection rate when the exponent of the power law of link number is below a critical value. 
 Several authors studied the contact process with quenched disorder~\cite{Noest, Cafiero}. In the disordered contact process below the critical value, there is a state called Griffiths phase, where the total infection decays to zero but the decay occurs very slowly. Griffiths proposed a mechanism of the slow decay originally in random spin systems~\cite{Griffiths}. 
There are compact infection clusters of $N$ sites whose lifetime is O$(e^{aN})$ in the random system, however, the probability of the clusters of $N$ sites is estimated as O$(e^{-bN})$.  As a result, the total infection decays as $t^{-b/a}$~\cite{Griffiths,Noest}. The Griffiths phases were studied in various complex networks.~\cite{Munoz, Moretti,Cota, Cota2} Even slower logarithmic decay was also reported in some complex networks.~\cite{Lee,Cota3}   
     
In this paper, we study the decay process of the infection in the deterministic Kermack-McKendrick model with spatially inhomogeneous infection rate in one, two, and three dimensions, considering that infection often spreads rapidly in densely populated zones or hot spots. 
The hot spots are inhomogeneously distributed. In this paper, we will show slow decay of infection such as a power-law decay with numerical simulations. 
In Sec.~I\hspace{-.1em}I, we make a brief review of the Kermack-McKendrick model as an ordinary differential equation. The SIR model has a unique property that there are infinitely large number of stationary solutions in the SIR model depending on the initial conditions. In Sec.~I\hspace{-.1em}I\hspace{-.1em}I, we study the Kermack-McKendrick model with one hot spot. We show the power-law decay of exponent $1/2$ in the one-dimension model with one hot spot. In Sec.~I\hspace{-.1em}V, we perform some numerical simulations of the Kermack-McKendrick model with quenched randomness in one, two, and three dimensions. In Sec.V, the results are summarized. Slow dynamics appears in various complex systems such as glassy soft matter~\cite{glass,glass2}. In the glassy soft matter, the dynamic heterogeneities of clusters and trapping in a random energy landscape are closely related to the slows dynamics~\cite{Adams}. The mechanism of slow dynamics in our model is different from that of the glassy state, in that the freezing of particle motions does not occur in our system.  
The slow decay in our model  might be related with the Griffiths phase, in that  quenched randomness is important in both systems. It is important that large clusters with large lifetimes    appear  with rare probabilities in the Griffiths phase of the contact process. On the other hand, the power-law decay of exponent $1/2$ in one dimension is due to the diffusion process as shown in Sec.~I\hspace{-.1em}I\hspace{-.1em}I in our model, and a coarsening process is observed as shown in Sec.~I\hspace{-.1em}V, which are not so important in the slow dynamics in the Griffiths phase. It is somewhat similar to the power-law decay found in the phase separation or ordering process near the first-order phase transition, where both the diffusion and coarsening processes are important~\cite{Gunton,Onuki}. We think that the diffusion, coarsening, and quenched randomness are important for the slow decay in our model, however, the theoretical understanding is not sufficient yet and the details are left to future study. 

\section{SIR model}
We make a very brief review of the dynamics of the Kermack-McKendrick model as an ordinary differential equation~\cite{KM,Murray}. 
The Kermack-McKendrick model is expressed as 
\begin{eqnarray}
\frac{dS}{dt}&=&-\beta SI,\\
\frac{dI}{dt}&=&\beta SI-\gamma I,
\end{eqnarray}
where $S$ and $I$ denote susceptible and infected populations. The parameters $\beta$ and $\gamma$ denote infection and recovering rates. The recovered population $R$ is calculated from 
\[\frac{dR}{dt}=\gamma I.\]
Since the three variables $S$, $I$, and $R$ are used, the Kermack-McKendrick model is called the SIR model. If the population $E$ of exposed persons before the appearance of symptoms is included, the SIR model is generalized to the SEIR model:\begin{eqnarray}
\frac{dS}{dt}&=&-\beta SI,\nonumber\\
\frac{dE}{dt}&=&\beta SI-\delta E,\nonumber\\
\frac{dI}{dt}&=&\delta E-\gamma I 
\end{eqnarray}
where $\delta$ denotes the incidence rate. In the SIR model, recovered persons are assumed not to be infected again owing to the immunity. However, there are diseases such as Malaria, for which there is a possibility that recovered persons are infected again. For such diseases, the SIS model where  $dS/dt=-\beta SI+\gamma I$ is used instead of Eq.~(1). In the SIR and SEIR models, $I$ decays to zero finally, but $I$ does not decay to zero in the SIS model. The contact process is a stochastic version of SIS model. 
In this paper, we consider mainly the SIR model, however, slow dynamics is observed even in the SEIR model as shown in Sec.~I\hspace{-.1em}V.  
\begin{figure}[b]
\begin{center}
\includegraphics[height=4.cm]{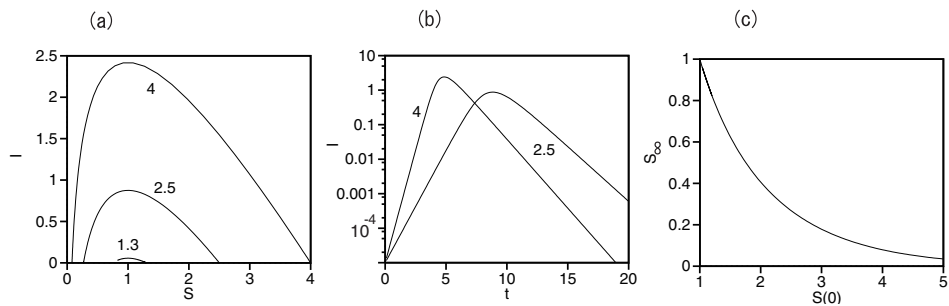}
\end{center}
\caption{(a) Trajectories in $(S,I)$ space starting from  $S(0)=4$, $2.5$, and $1.3$ at $\beta=\gamma=1$. (b) Time evolutions of $I(t)$ for $S(0)=4$ and 2.5 at $\beta=\gamma=1$ in the semi-logarithmic scale. (c) $S_{\infty}$ as a function of $S(0)$ at $\gamma=\beta=1$.}
\label{fig1}
\end{figure}

There is a stationary solution $S=S_0$ and $I=0$ to Eqs.~(1) and (2). The stationary state is unstable for $S_0\beta>\gamma$.  
Figure 1(a) shows trajectories in $(S,I)$ space starting from the initial condition  $S(0)=4$, $2.5$, and $1.3$ at $\beta=\gamma=1$. The initial condition for $I$ is fixed to be 0.00001. 
Figure 1(b) shows time evolutions of $I(t)$ for  $S(0)=4$ and 2.5 at $\beta=\gamma=1$ in the semi-logarithmic scale.
The infection $I(t)$ spreads initially and then decays to zero exponentially since the susceptible population $S(t)$ becomes below $\gamma/\beta$.  This is a state that the herd immunity is attained.  
The final state is $(S,I)=(S_{\infty},0)$ where $S_{\infty}<\gamma/\beta$ depends on the initial value $S(0)$. 
The uninfected population $S_{\infty}$ takes a smaller value for a larger initial value $S(0)$ of $S(t)$. $S_{\infty}$ can be calculated for the conserved quantity $Q$ of this equation:
\begin{equation}
Q=S+I-(\gamma/\beta) \log(S)
\end{equation}
If $I(0)$ is sufficiently small, $S_{\infty}$ is a solution of 
\[S_{\infty}-(\gamma/\beta) \log(S_{\infty})=S(0)-(\gamma/\beta)\log(S(0)).\]
Figure 1(c) shows $S_{\infty}$ as a function of $S(0)$ at $\gamma=\beta=1$. 
The ratio $S_{\infty}/S_0$ takes any value between 0 and 1, depending on the initial value $S_0$. 

\section{Slow decay of infection in SIR models with one hot spot}
Hereafter, we consider spatially extended systems. The spread of infection occurs in densely populated regions. In this paper, we consider mainly SIR models with inhomogeneous infection rate on one-, two-, ang three-dimensional lattices. In one dimension, the model equation is written: 
\begin{eqnarray}
\frac{dS_i}{dt}&=&-\beta_i S_iI_i+D_S(S_{i+1}-2S_i+S_{i-1}),\label{e1}\\
\frac{dI_i}{dt}&=&\beta_i S_iI_i-\gamma I_{i}+D_I(I_{i+1}-2I_i+I_{i-1}), \label{e2}
\end{eqnarray}
where $D_S$ and $D_I$ are diffusion constants. Periodic boundary conditions are imposed and the system size is $N$. 
The initial conditions were set to be $S_i=1$ and $I_i=0.00001$ for numerical simulations. 
Figure 2(a) shows time evolutions of the total population of infection: $SI=\sum_{i=1}^NI_i$ for $N=1000$ (solid line) and $N=200$ (dotted line) at $D_S=1,D_I=1,\gamma=1$ in the double-logarithmic scale. The infection rate is $\beta_i=\beta_o=0.9$ for $i\ne N/2$ and $\beta_i=\beta_p=3$ at $i=N/2$. The total infected population $SI$ decays exponentially for $N=200$, however, it decays slowly with a power law $0.4/t^{1/2}$ for $N=1000$.  
Even in the case of $N=200$, $SI$ decays in a power law of exponent $1/2$ until $t\sim 2000$. There is a crossover time from the power law decay to the exponential decay, and the crossover time increases with $N$.   
Figure 2(b) shows time evolutions of $SI=\sum_{i=1}^NI_i$ for $N=1000$ (solid line) and $N=200$ (dotted line) in the double-logarithmic scale, when the infection rate is $\beta_i=3$ for $N/2-7\le i\le N/2+7$ and $\beta_o=0.9$ for the other region. Other parameters are the same as in Fig.~2(a). 
The total infected population $SI$ decays with the power law of exponent $1/2$ for $N=1000$.  The damping oscillation is observed near $t=100$ in contrast to Fig.~2(a). 
\begin{figure}[h]
\begin{center}
\includegraphics[height=5.cm]{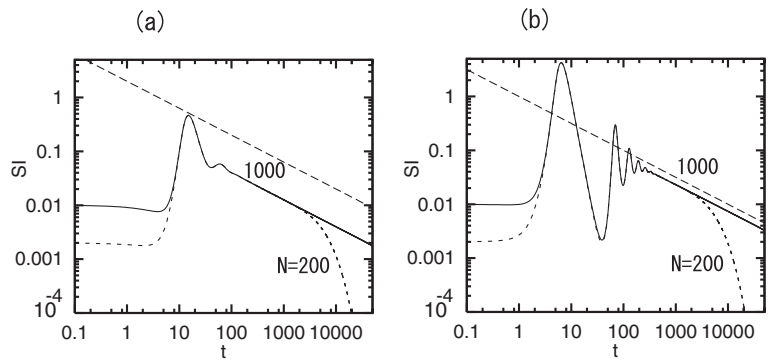}
\end{center}
\caption{(a) Time evolutions of $SI=\sum_{i=1}^NI_i$ for $N=1000$ (solid line) and $N=200$ (dotted line) at $D_S=1,D_I=1$, and $\gamma=1$ in the double-logarithmic scale. The infection rate is $\beta_i=\beta_o=0.9$ for $i\ne N/2$ and $\beta_i=3$ at $i=N/2$. The straight dashed line denotes a power law of $1/t^{1/2}$. (b) Time evolutions of $SI=\sum_{i=1}^NI_i$ for $N=1000$ (solid line) and $N=200$ (dotted line) in the double-logarithmic scale. The infection rate is $\beta_i=3$ for $N/2-7\le i\le N/2+7$ and $\beta_0=0.9$ for the other region. Other parameters are the same as in Fig.2(a).}
\label{fig2}
\end{figure}
\begin{figure}[h]
\begin{center}
\includegraphics[height=4.cm]{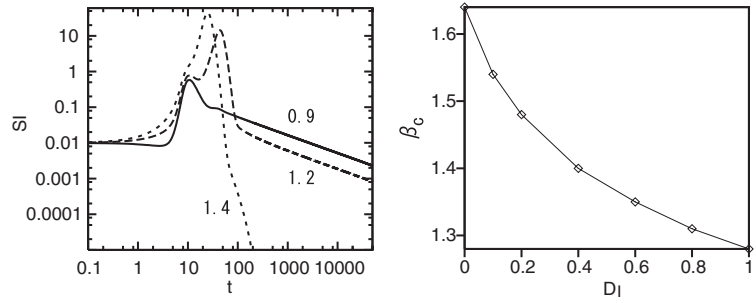}
\end{center}
\caption{(a) Time evolutions of $SI=\sum_{i=1}^NI_i$ at $\beta_o=0.9$ (solid line), (b) 1.2 (dotted line) and 1.4 (dotted line) for $N=1000$, $D_S=1,D_I=0.5,\gamma=1$ in the double-logarithmic scale.  (b) Critical value $\beta_c$ for the power-law decay as a function of $D_I$ for $N=1000$.}
\label{fig3}
\end{figure}

The power-law decay does not appear for large $\beta_o$, i.e.,the infection rate of $i\ne N/2$ under a fixed value of $\beta_i=\beta_p$ at $i=N/2$.  
Figure 3(a) shows the time evolutions of $SI=\sum_{i=1}^NI_i$ at $\beta_o=0.9$ (solid line), (b) 1.2 (dashed line) and 1.4 (dotted line) for $N=1000$. $D_S=1,D_I=0.5$, and $\gamma=1$ in the double-logarithmic scale. The power law of exponent $1/2$ is observed at $\beta_o=0.9$ and 1.2 but $SI$ decays rapidly at $\beta_o=1.4$.  Figure 3(b) shows the critical value of $\beta_o$ as a function of $D_I$ for $\gamma=1$ and $D_S=1$. In Fig.~2, we have shown numerical results for the case $\beta_oS_i(0)<\gamma$, however, the critical value is larger than $\gamma/S_i(0)=1$, that is, the power-law decay appears even when the infection spreads in the surrounding region. The critical value of $\beta_o$ decreases with $D_I$. The slow decay occurs in a wider parameter range for smaller $D_I$. 

Although the total infection decays to zero finally under the periodic boundary conditions, a stationary localized structure of infection is realized if the fixed boundary conditions $S_0=S_{N+1}=S_B$ are imposed. Similar localized structure is maintained for a long time even in systems of periodic boundary conditions. 
Figures 4(a) and (b) show stationary profiles of (a) $I_i$ and (b) $S_i$ for $N=200$ and $S_B=1$. The other parameters are  $D_S=1$, $D_I=1$, $\gamma=1$, and the infection rate is $\beta_i=0.9$ for $i\ne N/2$ and $\beta_i=3$ at $i=N/2$, which are the same as in Fig.~2(a). The infection occurs locally near $i=N/2$, $S_i$ diffuses into the infection region, and is infected near $i=N/2$. Outside of the infection region, the profile of $S_i$ has a nearly constant slope, that is, $S_i=S_B-\alpha i$ for $i<N/2$ and $S_i=S_B-\alpha (N-i)$ for $i>N/2$. The profile of $I_i$ is approximated as $I_i=I_0 e^{-\lambda |i-N/2|}$. Figure 4(c) shows the relationship between $\beta_{N/2}$ and $S_{N/2}$. The localized structure disappears for $\beta_{N/2}<1.5$ and $S_i=S_{N/2}=S_B=1$ is satisfied. 
\begin{figure}[h]
\begin{center}
\includegraphics[height=4.cm]{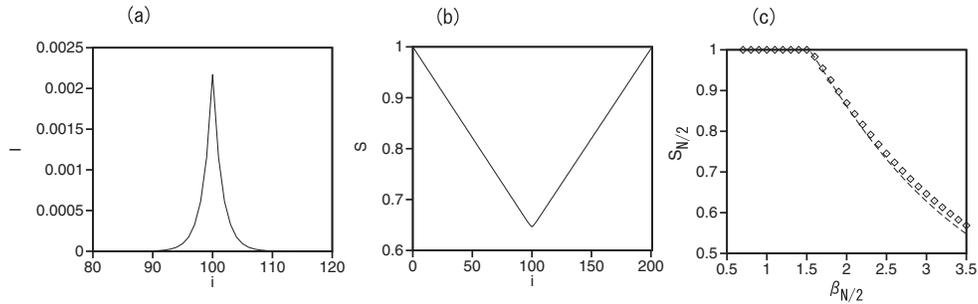}
\end{center}
\caption{Stationary profiles of (a) $I_i$ and (b) $S_i$ for $N=200$ and $S_B=1$. The other parameters are  $D_S=1,D_I=1,\gamma=1$, and the infection rate is $\beta_i=0.9$ for $i\ne N/2$ and $\beta_i=3$ for $i=N/2$. (c) Relationship between $\beta_{N/2}$ and $S_{N/2}$.}
\label{fig4}
\end{figure}

Next, we make an analysis of the localized state and the power-law decay of exponent 1/2.
If the continuum approximation is taken, Eqs.~(\ref{e1}) and (\ref{e2}) are rewritten as 
\begin{eqnarray}
\frac{\partial S}{\partial t}&=&-\beta SI+D_S\nabla^2S, \label{e3} \\
\frac{\partial I}{\partial t}&=&\beta SI-\gamma I+D_I\nabla^2I, \label{e4}
\end{eqnarray}
where $\nabla^2=\partial^2/\partial x^2$ in one dimension, however, the model will be later extended to two and three dimensions. 
For the stationary state in one dimension, $S(x)$ and $I(x)$ satisfy 
\begin{eqnarray}
D_S\frac{\partial^2S}{\partial x^2}&=&\beta(x)S(x)I(x),\\
D_I\frac{\partial^2I}{\partial x^2}&=&\{\gamma-\beta(x)S(x)\}I(x).
\end{eqnarray}
If $I(x)$ and $\beta(x)$ are assumed to be $I(x)=I(N/2) e^{-\lambda |x-N/2|}$ and $\beta(x)=\beta_0+\beta_1\delta(x-N/2)$, $\lambda$ satisfies
\begin{equation}
\lambda=\sqrt{(\gamma-\beta_0S(N/2))/D_I}.
\end{equation}
Equation (10) yields
\begin{equation}
D_I\left \{\left (\frac{\partial I}{\partial x}\right )_{(N/2)_{+}}-\left (\frac{\partial I}{\partial x}\right )_{(N/2)_{-}}\right \}=-\beta_1S(N/2)I(N/2)
\end{equation}
at $x=N/2$, which leads to  
\begin{equation}
2D_I\lambda=\beta_1S(N/2).
\end{equation}
Equations (11) and (13) yield
\begin{equation}
S(N/2)=\frac{\sqrt{4D_I^2\beta_0^2+4D_I\gamma \beta_1^2}-2D_I\beta_0}{\beta_1^2}.
\end{equation} 
The dashed line in Fig.~4(c) shows the relationship between $\beta_{N/2}=\beta_0+\beta_1$ and $S(N/2)$ by Eq.~(14). Fairly good agreement with direct numerical results is seen. 
If the infection region is sufficiently small, Eq.~(9) gives
\begin{equation}
D_S\left \{\left (\frac{\partial S}{\partial x}\right )_{(N/2)_{+}}-\left (\frac{\partial S}{\partial x}\right )_{(N/2)_{-}}\right \}=\beta_1S(N/2)I(N/2).
\end{equation}
From Eq.~(15), $I(N/2)$ is approximated at
\begin{equation}
I(N/2)=\frac{2D_S(S_B-S(N/2))/(N/2)}{\beta_1S(N/2)}
\end{equation}

If the boundary conditions are not fixed to $S_B$, the power-law decay is observed. Similar power-law decay is observed even for $D_I=0$. Figure 5(a) shows the time evolution of $I_{N/2}$ for $N=1000$ at $D_S=1,\;D_I=0$, and $\gamma=1$ in the double-logarithmic scale. The infection rate is $\beta_i=0.9$ for $i\ne N/2$ and $\beta_i=3$ at $i=N/2$. Figure 5(b) shows two snapshots of $S_i$ at $t=100$ and 5000. The population $S_i$ at $i=N/2$ is $1/3$ and the width of the depression increases as $t^{1/2}$. 
\begin{figure}[h]
\begin{center}
\includegraphics[height=4.cm]{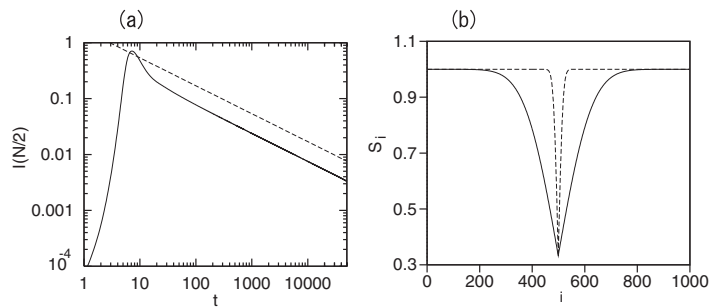}
\end{center}
\caption{(a) Time evolution of $I(N/2)$ for $N=1000$ at $D_S=1,\;D_I=0,\; \gamma=1$ in the double-logarithmic scale. The infection rate is $\beta_i=0.9$ for $i\ne N/2$ and $\beta_i=3$ for $i=N/2$. (b) Two snapshots of $S_i$ at $t=100$ and 5000.}
\label{fig5}
\end{figure}

In the continuum approximation, the solution satisfies
\begin{equation}
\frac{\partial S}{\partial t}=D_S\frac{\partial^2S}{\partial x^2}
\end{equation}
for $x\ne N/2$, since $I(x,t)$ is almost zero for $x\ne N/2$, and $2D_S\partial S/\partial x=\beta_{N/2}S(N/2)I(N/2)$ at $x=(N/2)_{+}$. Because $S(N/2)$ is fixed to be $\gamma/\beta_{N/2}$ in case of $D_I=0$, 
$I(N/2)$ is determined to be $2D_S\partial S/\partial x/\gamma$.  That is, the population of infection is determined by the diffusion process of susceptible population for large $t$. 
The slope $\partial S/\partial x$ at $x=(N/2)_{+}$ can be calculated from the solution of $S(x,t)$ to the diffusion equation. 
The diffusion equation can be solved by the Fourier transform 
\begin{equation}
S(x,t)=S(N/2)+\frac{1-S(N/2)}{N/2}(x-N/2)+\sum_{n=1}^{\infty}A_n(t)\sin\{2\pi n (x-N/2)/N\},\end{equation}
for $x>N/2$. The Fourier amplitude $A_n(t)$ satisfies $A_n(t)=A_n(0)e^{-D_S4\pi^2 n^2t/N^2}$, where $A_n(0)$ is determined from the initial condition $S(x,0)=(S(N/2)-1)\delta (x-N/2)+1$ as  
\[A_{n}(0)=(1-S(N/2))\frac{2}{\pi n}.\]
$I(N/2)$ is evaluated as
\[I(N/2)=2D_S\frac{\beta_{N/2}}{\gamma}\frac{\partial S}{\partial x}=2D_S\frac{\beta_{N/2}}{\gamma}\sum_{n=1}^{\infty}(1-S(N/2))\frac{2}{N}e^{-4D_S\pi^2n^2t/N^2}.\]
Since the summation is negligible for $n$ satisfying $4D_S\pi^2n^2t/N^2>>1$ or $n>>N/(4D_S\pi^2 t)^{1/2}$, $I(N/2)$ is approximated as
\[I(N/2)=\frac{4(1-\gamma/\beta_{N/2})\sqrt{D_s}}{\pi\gamma\sqrt{t}},\]
if the contribution $(1-S(N/2))/(N/2)$ in the second term of Eq.~(18) is neglected for large $N/2$. 
This is a reason of the power law of exponent $1/2$. The dashed line in Fig.~5(a) denotes this relation, which is good approximation to the direct numerical simulation.

Two- and three dimensional models are expressed with Eqs.~(\ref{e3}) and (\ref{e4}) if $\nabla^2$ is rewritten as $\partial^2/\partial x^2+\partial^2/\partial y^2$ in two dimensions and  $\partial^2/\partial x^2+\partial^2/\partial y^2+\partial^2/\partial z^2$ in three dimensions.
If $S$, $I$ and $\beta$ depends only the radius $r$, Eqs.~(\ref{e3}) and (\ref{e4}) are expressed as 
\begin{eqnarray}
\frac{\partial S}{\partial t}&=&-\beta SI+D_S\left\{\frac{\partial^2S}{\partial r^2}+\frac{d-1}{r}\frac{\partial S}{\partial r}\right \},  \label{e5}\\
\frac{\partial I}{\partial t}&=&\beta SI-\gamma I+D_I\left\{\frac{\partial^2I}{\partial r^2}+\frac{d-1}{r}\frac{\partial I}{\partial r}\right \}, \label{e6} 
\end{eqnarray}
where $d$ denotes the dimension 2 or 3. 
We have performed numerical simulation of Eqs.~(\ref{e5}) and (\ref{e6}) as a one-dimensional discrete system similar to Eqs.~(\ref{e1}) and (\ref{e2}) with additional terms of $D_S(d-1)/r\partial S/\partial r$ and $D_I(d-1)/r\partial I/\partial r$. 
The system size is  $N=1000$, and $r=0$ is set in the middle of  $i=N/2$ and $N/2+1$. That is, $r=i-(N+1)/2$.  
Parameters are $D_S$, $\gamma=1$, and $\beta_i=0.9$ for $i\ne N/2, N/2+1$ and $\beta_i=3$ for $i=N/2$ and $N/2+1$. Figure 6(a) shows the time evolution of $I(N/2)$ at $D_I=1$ (solid line) and $D_I=0$ (dashed line) for $d=2$. Figure 6(b) shows $1/I(N/2)$ as a function of $\log(t)$ at $D_I=1$ (solid line) and $D_I=0$ (dashed line). The solid line can be approximated at $5.8\log(t)+1.2$ and the dashed at $0.71\log(t)+0.5$. That is, $I(N/2)$ decays as $1/\log(t)$, The decay of infection in two dimensions is slower than in one dimension. 
Figure 6(c) shows a snapshot of $S(x,t)$ at $t=5000$ for $D_I=0$. The value of $S(r)$ is almost equal to be $\gamma/\beta(N/2)=1/3$, however, the slope of $S(r)$ seems to be divergent at $r=0$ in contrast to the one-dimensional case.  
The analysis using the expansion by the Bessel and Neumann functions instead of the Fourier series expansion might explain the logarithmic decay.   
Figure 6(d) shows the time evolution of $I(N/2)$ at $D_I=1$ (solid line) and $D_I=0$ (dashed line) for $d=3$. The other parameter values are the same as in Fig.~6(a). The decay of $I(N/2)$ is even slower in three dimensions. 
The functional form of the decay is not known yet. 
The infection hardly disappears, probably because the susceptible population in the surrounding region diffuses into the center of the three-dimensional hot spot and the site number of the surrounding region is large in the three-dimensional system compared to the one- and two-dimensional systems.    
\begin{figure}[h]
\begin{center}
\includegraphics[height=4.cm]{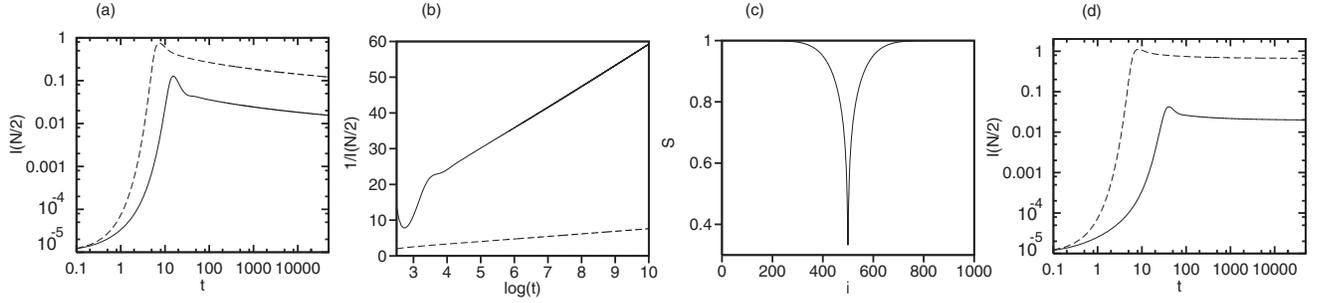}
\end{center}
\caption{(a) Time evolution of $I(N/2)$ at $D_I=1$ (solid line) and $D_I=0$ (dashed line) for $d=2$. Parameters are $D_S=1$, $\gamma=1$, and $\beta_i=0.9$ for $i\ne N/2, N/2+1$ and $\beta_i=3$ for $i=N/2$ and $N/2+1$. (b) $1/I(N/2)$ as a function of $\log(t)$ at $D_I=1$ (solid line) and $D_I=0$ (dashed line). (c) Snapshot profile of $S(x,t)$ at $t=5000$ for $D_I=0$. (d) Time evolution of $I(N/2)$ at $D_I=1$ (solid line) and $D_I=0$ (dashed line) for $d=3$. }
\label{fig6}
\end{figure}

\section{Slow decay of infection in random SIR models}  
In this section, we study random SIR models in one, two and three dimensions. 
The one-dimensional model is again expressed as 
\begin{eqnarray}
\frac{dS_i}{dt}&=&-\beta_i S_iI_i+D_S(S_{i+1}-2S_i+S_{i-1}),\nonumber\\
\frac{dI_i}{dt}&=&\beta_i S_iI_i-\gamma I_{i}+D_I(I_{i+1}-2I_i+I_{i-1}). \label{e7}
\end{eqnarray}
where $\beta_i$ is a uniform random number between 0 and $\beta_m$. 
Figure 7(a) shows time evolutions of $SI=\sum_{i}I_{i}$  for Eq.~(\ref{e7}) at $D_I=1$ and $\beta_m=1.5$ (solid line) and $D_I=0$ and $\beta_m=1.1$ (dashed line). The other parameters are $D_S=1$ and $\gamma=1$.  The system size is $N=5000$. The initial condition is $S_i=1$ and $I_i=0.00001$. Figure 7(a) shows that $SI$ decays roughly with a power law for large $t$. The exponent is around 1 for $D_I=1$ and $\beta_m=1.5$, and 1.35 for $D_I=0$ and $\beta_m=1.1$. The exponent depends on the parameters,   
 Figure 7(b) shows three snapshot profiles of $I_i$ at $t=200,300$ and 2000 for $D_I=1$ and $\beta_m=1.5$. The other parameters are the same as the ones in Fig.~7(a). Figure 7(c) shows four snapshot profiles of $S_i$ at $t=200,300,2000$, and 10000 in the same frame. The infection occurs at many hot spots at $t=200$, and the spatial profile is intermittent as shown in Fig.~7(b). The number of hot spots decrease with time. The cusp points in Fig,~7(c) correspond the hot spots. $S_i$ decreases with time by the diffusion to the hot spots and infection at the hot spots. Figure 7(d) shows the local average of $\beta_i$ (red dotted line), that is, $\bar{\beta}_i=(1/2)\sum_{j=i-10}^{i+10}\beta_i$, $100I_i$ (blue dashed line) at $t=1000$ and the maximum value of $I_i(t)$ (green solid  line) for $0<t<10000$ in the range of $2000<i<3000$. 
In most cases, hot spots appear near the points where the locally averaged infection rate $\bar{\beta}_i$ is large. Infection is stamped out at some hot spots, however, strong hot spots survive for long. The lifetime of a hot spot with an interval $L$ is estimated as O$(L^2)$, because the width of the diffusion field increases as $t^{1/2}$ as shown in Fig.~5(b) and the power-law decay changes to an exponential decay when the width reaches the size of interval. If the intervals between strong hot spots become longer after the burnout of some hot spots, the lifetime of the survived hot spot becomes even longer.  The power-law decay of $t^{-1/2}$ by the diffusion effect and the increase of lifetime by the coarsening process might be the origin of the slow decay in one dimension. 
\begin{figure}[h]
\begin{center}
\includegraphics[height=3.7cm]{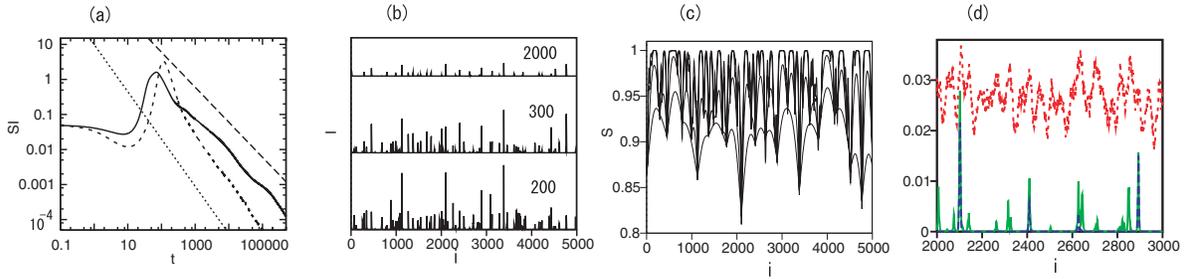}
\end{center}
\caption{(a) Time evolutions of $SI=\sum_{i}I_{i}$  for Eq.~(\ref{e7}) at $D_I=1$ and $\beta_m=1.5$ (solid line) and $D_I=0$ and $\beta_m=1.1$ (dashed line) for $D_S=1$,and $\gamma=1$. (b) Three snapshot profiles of $I_i$ at $t=200,300$ and 2000 for $D_I=1$ and $\beta_m1.5$. (c) Four snapshot profiles of $S_i$ at $t=200,300,2000$, and 10000 from above. (d) Local average of $\beta_i$ (red dotted line), $100I_i$ (blue dashed line) at $t=1000$ and the maximum value of $I_i(t)$ for $0<t<10000$ (green solid line).}
\label{fig7}
\end{figure}

We study the SEIR model to check the generality of the slow dynamics: The model equation is  
\begin{eqnarray}
\frac{dS_i}{dt}&=&-\beta_i S_iI_i+D_S(S_{i+1}-2S_i+S_{i-1}),\nonumber\\
\frac{dE_i}{dt}&=&\beta_i S_iI_i-\delta_i E_i+D_E(E_{i+1}-2E_i+E_{i-1}),\nonumber\\
\frac{dI_i}{dt}&=&\delta_i E_i-\gamma I_{i}+D_I(I_{i+1}-2I_i+I_{i-1}), \label{se2}
\end{eqnarray}
where $S_i$, $E_i$ and $I_i$ denote respectively the susceptible, exposed, and infected populations, and $\delta_i$ denotes the incidence rate. 
The parameters are set to be $D_S=D_E=1$, $D_I=0.5$, $\gamma=1$, and $\delta=2$. 
Figure 8(a) shows the time evolutions of $SI=\sum I_i$ in a system with one hot spot for $N=1000$, where $\beta_i=\beta_o=0.9$ (solid line), 1.2 (dashed line), and 1.4 (dotted line) for $i\ne N/2$ and $\beta_i=3$ at $i=N/2$. 
The power law decay of exponent $1/2$ is observed for $\beta_o=0.9$ and 1.2. An exponential decay is observed at $\beta_0=1.4$. 
Figure 8(b) shows the time evolution of $SI$ in a random SEIR model of $N=5000$ where $\beta_i$ takes a uniform random number between 0 and 1.4. $SI$ exhibits a power-law decay of exponent around 0.85. These results are similar to those in the SIR model. 
\begin{figure}[h]
\begin{center}
\includegraphics[height=4.cm]{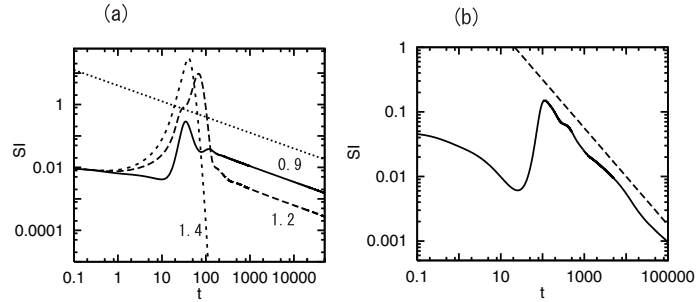}
\end{center}
\caption{(a) Time evolutions of $SI=\sum I_i$ in a system with one hot spot for $\beta_o=0.9$ (solid line), 1.2 (dashed line), and 1.4 (dotted line) for $N=1000$, $D_S=D_E=1$, $D_I=0.5$, $\gamma=1$, and $\delta=2$. (b) Time evolution of $SI$ in a random SEIR model of $N=5000$ where $\beta_i$ takes a uniform random number between 0 and 1.4. The dashed line denotes a power law of exponent 0.85.}
\label{fig8}
\end{figure}

The two-dimensional random SIR model is expressed as 
\begin{eqnarray}
\frac{dS_{i,j}}{dt}&=&-\beta_{i,j} S_{i,j}I_{i,j}+D_S(S_{i+1,j}+S_{i-1,j}+S_{i,j}+S_{i,j-1}-4S_{i,j}),\nonumber\\
\frac{dI_{i,j}}{dt}&=&\beta_{i,j} S_{i,j}I_{i,j}-\gamma I_{i,j}+D_I(I_{i+1,j}+I_{i-1,j}+I_{i,j+1}+I_{i,j-1}-4I_{i,j}), \label{e8}
\end{eqnarray} 
Figure 9(a) shows time evolutions of $SI=\sum_{i,j}I_{i,j}$  for Eq.~(\ref{e8}) at $D_S=D_I=1, \gamma=2$ in random systems, when $\beta_{i,j}$ takes a uniform random value between 0 and $\beta_m=3.5$ (solid line) or between 0 and $\beta_m=3.2$ (dashed line). The dotted line is $SI\propto 1/t^{0.85}$. The system size is $600\times 600$. The initial condition is $S_{i,j}=1$ and $I_{i,j}=0.00001$. The total population of infection seems to decay in a power law in these numerical simulations.
Figure 9(b) shows some snapshots of $I_{i,j}$ at a section of $j=N/2$ when $\beta_{i,j}$ takes a uniform random value between 0 and 3.2 at $t=50,100,\cdots$, 500. Localized clusters of infection survive for long. The number of hot spots decreases with time, or a coarsening occurs also in two dimensions. 
\begin{figure}[h]
\begin{center}
\includegraphics[height=4.5cm]{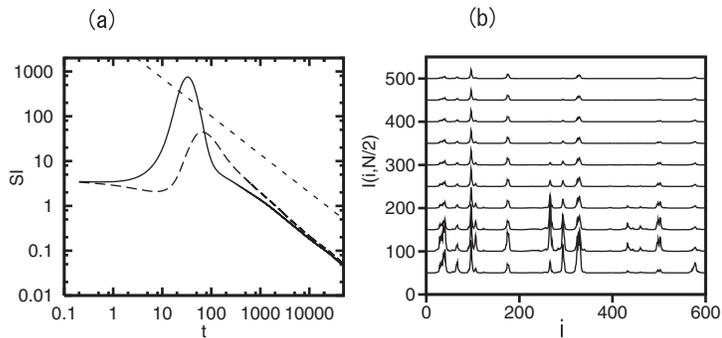}
\end{center}
\caption{(a) Time evolutions of $SI=\sum_{i,j}I_{i,j}$  for Eq.~(\ref{e8}) at $D_S=D_I=1, \gamma=2$ and $\beta_m=3.5$(solid line) and $\beta_m=3.2$ (dashed line). The dotted line is $SI\propto 1/t^{0.85}$. The system size is $600\times 600$. (b) Snapshots of $I_{i,j}$ at a section of $j=N/2$ when $\beta_{i,j}$ for $\beta_m=3.2$ at $t=50,100,\cdots$, 500.}
\label{fig9}
\end{figure}

The three-dimensional model is expressed as
\begin{eqnarray}
\frac{dS_{i,j,k}}{dt}&=&-\beta_{i,j,k}S_{i,j,k}I_{i,j,k}+D_S(S_{i+1,j,k}+S_{i-1,j,k}+S_{i,j+1,k}+S_{i,j-1,k}+S_{i,j,k+1}+S_{i,j,k-1}-6S_{i,j,k}),\nonumber\\
\frac{dI_{i,j,k}}{dt}&=&\beta_{i,j,k} S_{i,j,k}I_{i,j,k}-\gamma I_{i,j,k}+D_I(I_{i+1,j,k}+I_{i-1,j,k}+I_{i,j+1,k}+I_{i,j-1,k}+I_{i,j,k+1}+I_{i,j,k-1}-6I_{i,j,k}), \label{e9}
\end{eqnarray} 
Figure 10(a) shows time evolution of $I_{N/2+1,N/2+1,N/2}$ for Eq.~(\ref{e9}) at $D_S=D_I=0.2$, and $\gamma=1$ when $\beta_{i,j,k}=3$ for $i=N/2,N/2+1$, $j=N/2,N/2+1$, $k=N/2,N/2+1$ and $\beta_{i,j,k}=0.9$ for the other sites.  The system size is $100\times 100\times 100$. The initial condition is $S_{i,j,k}=1$ and $I_{i,j,k}=0.00001$. This is a three-dimensional simulation similar to the case shown in Fig.~6(d). $I_{N/2+1,N/2+1,N/2}$ is almost constant after $t>50$, that is, the localized infection is maintained for very long. Figure 10(b) shows time evolutions of $SI=\sum_{i,j,k}I_{i,j,k}$  for Eq.~(\ref{e9}) at $D_S=D_I=0.2$, and $\gamma=5$ in random systems, where $\beta_{i,j,k}$ takes a uniform random value between 0 and 10 (solid line) or between 0 and 6 (dashed line). The dotted line is $SI\propto 1/t^{1.1}$. The total population of infection seems to decay in a power law also in these random three-dimensional systems.
\begin{figure}[h]
\begin{center}
\includegraphics[height=4.5cm]{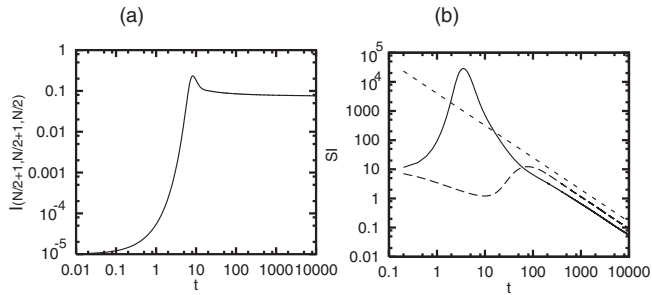}
\end{center}
\caption{ (a) Time evolution of $I_{N/2+1,N/2+1,N/2}$ for Eq.~(\ref{e9}) at $D_S=D_I=0.2$, and $\gamma=1$ when $\beta_{i,j,k}=3$ for $i=N/2,N/2+1$, $j=N/2,N/2+1$, $k=N/2,N/2+1$ and $\beta_{i,j,k}=0.9$ for the other sites.  The system size is $100\times 100\times 100$. (b) Time evolutions of $SI=\sum_{i,j,k}I_{i,j,k}$  for Eq.~(\ref{e9}) at $D_S=D_I=0.2$, and $\gamma=5$ in random systems, where $\beta_{i,j,k}$ takes a uniform random value between 0 and 10 (solid line) or between 0 and 6 (dashed line). The dotted line is $SI\propto 1/t^{1.1}$. }
\label{fig5}
\end{figure}
\section{Summary}
We have found slow decay of infection in the Kermack-McKendrick model with spatially inhomogeneous infection rate in some parameter range. First, we have studied the Kermack-McKendrick model with one hot spot where the infection rate is locally higher than the surrounding region. We have shown theoretically a power-law decay of exponent $1/t^{1/2}$ in the one-dimensional system with a spatially localized hot shot. The slow decay occurs as $1/\log t$ in the two-dimensional system with one hot spot, and the decay seems to be even slower in three dimensions. 
Next, we have studied the random Kermack-McKendrick model, and found the power-law type slow decay in one, two, and three dimensions. 
We found that the infection occurs locally at hot spots. The uninfected persons in the surrounding area around the hot spots diffuse into the hot spots and are infected at the hot spots. The number of hot spots decreases in time and the lifetime of the survived hot spots become even longer because the surrounding areas of the survived hot spots become larger. The mechanism of the slow decay in our system is unique in that the diffusion, coarsening,  and quenched randomness are important, although there is some similarity to the slows dynamics in the Griffiths phase in the contact process~\cite{Noest,Cafiero} and the phase transition dynamics~\cite{Gunton,Onuki}.  However, the slower decay in our two- and three- dimensional SIR models with one hot spot, and the exponent of the power-law decay in one-, two-, and three-dimensional random SIR models are not well understood. They are left to future study. Our finding of the slow decay of the infected population suggests that the infection is locally maintained for long time and the diseases are hardly stamped out in some cases.

\end{document}